# Use of Service Curve for Resource Reservation in Wired-cum-Wireless Scenario

Nitul Dutta
Sikkim Manipal Institute of Technology, Computer Science & Engg. Deptt., India
.

Iti Saha Misra
Jadavpur University, Electronics & Telecommunication Engineering Department, India
.

*Abstract*- *In a network, arrival process is converted into departure process through network elements. The departure process suffer propagation delay in the link, processing delay at the network elements like router and data loss due to buffer overflow or congestion. For providing guaranteed service resources need to be reserved before conversation takes place. To reserve such resources estimation of them are indispensable. The idea of service curve gives beforehand deterministic value of these parameters. In this paper, we aim to minimum and maximum buffer space required in the router, minimum link capacity required to guarantee a pre-specified end-to-end delay for an ongoing session in a wired-cum-wireless scenario by analyzing minimum and maximum service curve. We assume that the network we are analyzing is an IP based mobile network. The findings of the work are presented in the form of tables which can be used for resource reservation to offer quality service to end-users.*

I. INTRODUCTION

Proper analysis using network engineering techniques enhance the strength of any network architecture for world wide deployment. This analysis may be done either by probabilistic or deterministic method. In this work we want to study the behavior of traffic flow in a mobile IP based wired cum wireless scenario by means of deterministic method for an intrinsically stochastic stream source to provide guaranteed service. Network calculus has undergone a lot of development as a sophisticated tool for performance analysis in traffic engineering. This tool can be used to engineering networks with worst case performance guarantees in presence of variety of network elements like shapers, multiplexer and propagation delay. Here, emphasis is given to evaluate the performance of the network taking into account the minimum and maximum service curve so that a measure of minimum and maximum end-to-end delay, requirement of buffer and link capacity for a wired-cum wireless scenario can be made in advance to provide quality of service (QoS) guarantee in the network. Minimum service curve gives a measure of minimum data required at a network element (say router) for efficient utilization of the service rate and the upper service curve provides a measure of amount of data which may be handled by different network elements preserving the QoS to the ongoing traffic.

Rest of the paper is organized as follows. There are a lot of research work carried out on network calculus and its application in performance measure of network architecture. Section II is a discussion of few of such works with the motivation. Mathematical overviews of the network calculus along with a few definitions are found in section III which will help readers to have a clear understanding of the rest of the paper. The concept of service curve and its application to measure various resource requirements for providing guaranteed service is elaborated in section IV. The wired-cum-wireless scenario for which we have carried out the various service curve analysis is briefly explained in section V. Findings of this work are presented in a tabulated form in section VI along with a brief discussion of various parameters used for calculation. Finally, paper is concluded in section VII.

II. RELATED WORK AND MOTIVATION

A concise discussion on some of the research work related to network calculus and its application is presented in this section. The work of [1] presents a non-probabilistic approach for bursty data is modeled for deterministic bound through network calculus. The calculus they have developed gives a measure of bounds on delay and buffer requirement in a packet switched network under a fixed routing strategy. Another similar research presented in [2], have analyzed a conceptual framework for flow of data in integrated services network model and analyzed the behavior of open loop, rate based flow control protocols, as well as closed loop, window based flow control protocols. They have modeled links, propagation delays, schedulers, regulators, and window based throttles by using lower and upper service curve. Finally, a measure of end-to-end delay bounds and maximum buffer requirement is measured using service curves. The paper provides a good understanding of service curve for performance analysis with examples. Lot of propositions are proposed and proved in the paper which may be used as a basis for other research work. The objective of [3, 4] is to introduce stochastic network calculus as an evolving new

This work is supported by All India Council For Technical Education (AICTE), New Delhi, India under Research Promotion Scheme (RPS) F.No. 8032/BOR/RID/RPS-234/2008-09.
.

methodology for backlog and delay analysis of networks that can account for statistical multiplexing gain. They promote the advances of stochastic network calculus by deriving a network service curve, which expresses the service given to a flow by the network as a whole in terms of a probabilistic bound. Through the presented network service curve a calculation of statistical end-to-end delay and backlog bounds for broad classes of arrival and service distributions are measured. The benefits of the derived service curve are illustrated for the exponentially bounded burstiness (EBB) traffic model. Their work deals with the traffic flow through a series of network elements from source to destination and shows that end-to-end performance measures computed through a network service curve are bounded by O(HlogH), where H is the number of network elements traversed by the flow. Using currently available techniques, which compute end-to-end bounds by adding single node results, the corresponding performance measures are bounded by O($H_3$). Their work is inclined towards statistical approach rather than deterministic approach, which describes arrival process and offered service probabilistically while preserving the elegance and expressiveness of the original framework.

In wired-cum-wireless networks, resources in the wireless part is more precious as compared to that in the wired part. The situation becomes more complicated when wireless nodes are mobile. There are plenty of network layer protocols to support communication efficiency in such wireless environment. Mobile IPv6 (MIPv6) [5] and Hierarchical mobile IPv6 (HMIPv6) [6] are two widely used protocols with seamless mobility and best effort service. But in most of the cases guaranteed service is quite demanding rather than a best effort service. Many authors have proposed the use of Integrated Services (IntServ) [7] with Resource Reservation Protocol (RSVP) [8, 9, 10] to provide guaranteed service when MIPv6 or HMIPv6 protocols are used. But while using RSVP, measurement of resource requirement for any traffic flow is a complicated task. Especially in wireless environment the situation is worst. In such a situation network calculus is supposed to be the best method to estimate the amount of resource requirement for guaranteed service under various traffic conditions. This fact motivated us to do this research in a mobile IPv6 based wired-cum wireless scenario to make use of RSVP protocol.

III. MATHEMATICAL MODELING

Service curve specifies how the arriving stream of packets is converting into a departing process. Service curve can model a variety of network elements like router, link capacity etc. Before we proceed further, some definitions are presented here for clarity of the reader. These definitions are used in the subsequent sections in the paper.

*Definition 1. Process:* A process $A(t), t \in \Re$ is defined as a non-decreasing and right continuous function of time which could count the amount of data arriving or departing to or from some network element [12,13].

$A(t)$ is a mapping from the real number in to the extended nonnegative real numbers which is mathematically represented as $A : \Re \to \Re_+ \cup \{+\infty\}$

*Definition 2. Causal process:* A process $A(t), t \in \Re$ is said to be causal if it has the property $A(t) = 0$ for t<0 [12].

A causal process will be identically zero for all negative times. For example if $A(t)$ is a causal process representing the amount of traffic fed to an agent, then $A(t)$ is the amount of data (in bits) arriving to the agent in the interval $(-\infty, t]$.

*Definition 3. Supremum:* Let S is a subset of real number $\Re$ such that $\exists b \in \Re$, and $\forall s \in S, s \leq b$. Then subset S is said to be bounded above. In such case it can be proved that there is a number u and i) $\forall s \in S, s \leq u$ and ii) For every b, b upper bounds S, $u \leq b$. The number u is the least upper bound of S and called supremum or sup of set S. It is denoted as $u = \sup\{s : s \in S\}$ [12].

Supremum is the smallest real number that is greater than or equal to every number in S. A real number is said to have the completeness property and for this property every non-empty subset of real number that is bounded above has a supremum that is also a real number.

*Definition 4. Infimum:* Let S is a subset of real number $\Re$ such that $\exists a \in \Re$, and $\forall a \in S, a \leq s$. Then subset S is said to be bounded below. Then there is a greatest lower bound l which is called the infimum of S and is denoted by $l = \inf\{s : s \in S\}$. [12].

Infimum is the biggest real number that is smaller or equal to every real number in S. If no such number exists, then inf(S) = -∞ and if S is empty then inf(s) = ∞.

*Definition 5. Convolution operation (*):* If $A(t)$ and $B(t), t \in \Re$, are non negative non decreasing causal functions then, convolution of these two functions is defined as $(A * B)(t) := \inf_{\tau \in \Re}(A(\tau) + B(\tau - t))$ [4,12].

Convolution operation produces a third function from any two functions and may be viewed as modified version of one of the original functions.

*Definition 6. Identity element* $\delta(t)$ is defined as a process such that $\delta(t) = 0$ for t<0 and ∞ otherwise and satisfies the equality $\delta_d(t) = \delta(t - d)$ [12]. For any causal process A(t), $(A * \delta_d)(t) = A(t)$

## IV. SERVICE CURVE CONCEPT

Service curve scientifically and mathematically justifies the theoretical utilization and efficiency based on the incoming traffic to any network system. Service curve may be either lower or upper. The lower service curve deterministically evaluates amount of network traffic which is must to proper utilization of the service of the multiplexer. On the other side, the upper service curves gives a deterministic measure of maximum allowable data which could be fed to the multiplexer without degrading the end-to-end QoS. Both the upper and the lower service curve are dependent on the traffic arrival rate, service rate or processing rate of the multiplexer and bandwidth of the outgoing line in the multiplexer. To understand the concept clearly, let us consider Figure 1.

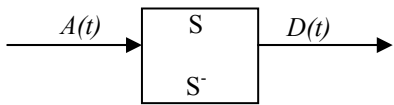

Figure 1: Service curve of a network element

In a network multiplexer, *A(t)* is the amount of traffic arrived at a duration of time *t* which is passed through the link with the capacity *C* bps and *D(t)* is the amount of data which is being departed within time *t*. Let us assume that the source data generates in minimum *r* bytes per sec and delivers to the multiplexer as a packet of $L_{max}$ bytes which is transmitted through the multiplexer. If data is not regulated by the source before delivering it to the channel for transmission, then any byte may be delayed by at most $\frac{L_{max}}{r}$ unit time. At time $t - \frac{L_{max}}{r}$, arrived traffic is less than the traffic that could be departed from the system. So,

$$D(t) \geq A(t - \frac{L_{max}}{r}) \quad (1)$$

From the definition of identity element δ(t) and convolution operation of identity element with A(t), we can rewrite equation (1) as

$$A(t - \frac{L_{max}}{r}) = (A * \delta_{\frac{L_{max}}{r}})(t)$$

i.e

$$D(t) \geq (A * \delta_{\frac{L_{max}}{r}})(t) \quad (2)$$

Hence, $\frac{\partial_{L_{max}}}{r}$ is the lower service curve. If $\frac{\partial_{L_{max}}}{r}$ is very small, service rate of the multiplexer is under-utilized. Again, if the source generates data in such a speed that before time $t - \frac{L_{max}}{r}$ the amount of data exceeds the data that the multiplexer could server, hence backlog will generate.

Let us now define lower and upper service curves mathematically. Let S*(t)* and $\bar{S}(t)$ are two non-negative, non-decreasing causal functions. If $D(t) \geq A(t) * S(t)$, then *S(t)* is a lower service curve and if $D(t) \leq A(t) * \bar{S}(t)$, then $\bar{S}(t)$ is called upper service curve. The operator * is the convolution operation. If $S(t) = \bar{S}(t)$ then, *S* is called service curve[12].

## V. THE NETWORK SCENARIO

In Figure 2, a sample network scenario is depicted for which various performance parameters are analyzed in accordance with the service curve. Different components of the scenario are discussed below.

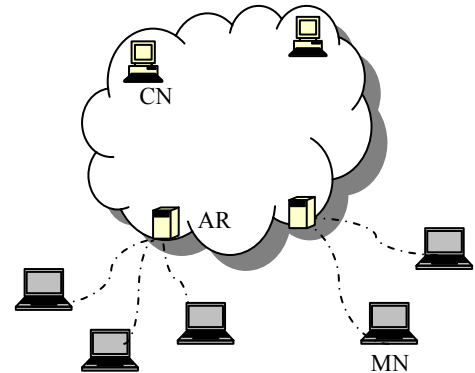

Figure 2: Example Wired-cum-Wireless scenario

A mobile node (MN) is connected to Access Router (AR). AR serves MNs within a specified region called cell. All MNs in the cell communicate with the Correspondent Node (CN) via same AR. We are intended to study delay suffered by packets from MN to CN, buffer space required at AR to store access data so that guaranteed service can be offered by reserving resources beforehand. A measure of minimum bandwidth required to support guaranteed service with a pre specified delay is also made in the analysis. Observation is made by splitting the entire scenario into two parts; wireless and wired segment. First we calculate the minimum and maximum delay of packets from MN to AR and then from AR to CN. We assume that several audio sessions are connected to CN from MN and traffic to wireless channel is injected by an envelope of (σ, ρ). There is no such regulator maintained at AR and data is allowed to pass to the wired link as long as bandwidth is available and buffered otherwise. A measure of minimum bandwidth required to maintain a pre specified end-to-end delay is also made for the scenario.

Under an AR there may be a group of MNs. The arrival process *A(t)* is influenced by the traffic of MNs under the coverage of the AR. As the number of MNs which are distributed under different ARs increases, the *A(t)* increases. The service rate of the AR remains constant. This is also true for departure process *D(t)* which may vary depending upon *A(t)* as long as it does not exceed the total capacity of the outgoing line of the AR. Once the arrival process exceeds the capacity of the outgoing link or the service rate of the AR is not sufficient to attend the arrival process, backlog is generated. A fixed length buffer is assumed to be maintained in each of the AR to minimize data loss. Hence, if the buffer is full the data will be dropped and performance of the network is said to be degraded. As a result, to maintain desired quality of services to the network traffic, either we have to restrict the total number of MNs under an AR or session should be stopped at the connection establishment phase.

In the next subsection we will discuss delay, buffer space and link capacity with respect to the above network scenario so that a measure could be made to provide preferential treatment to certain traffic in the network.

*A. Delay analysis*

Delay experienced by traffic passing through any network element is either upper and lower bounded by $d_{max}$ and $d_{min}$ respectively. If delay exceeds $d_{max}$ then the packet is of no use to the end. If $d_{min}$ is very low and end user cannot consume the data at the rate with which it arrives, buffer overflow will occur in the destination. If more packets are lost due to buffer overflow, end system may not make use of the arrived data properly. To support guaranteed service to end users, the values of $d_{max}$ and $d_{min}$ should be within the range such that end users can tolerate the delay or can make the other alternative (like buffer) to adjust the delay. Again, let us examine the delay bound in a traffic flow with arrival process *A(t)* and departure process *D(t)*. From the definition of *A(t)* and *D(t)* and value of $d_{min}$ and $d_{max}$ we can have

$$A(t - d_{max}) \leq D(t) \leq A(t - d_{min}) \quad (3)$$

It implies that amount of packets arrived till the time *t-$d_{max}$* cannot exceed the amount of traffic departed from the system but the amount of traffic arrived at time *t-$d_{min}$* may exceed the amount departed from the system at time *t*.

To provide guaranteed service to end users these two parameters must be selected carefully taking both the characteristics of the *A(t)* and *D(t)* into account. Using the shift operator and its definition $\delta_d(t) = \delta(t-d)$ where, $\delta_d$ is the shift operator and for any process $A*\delta_d(t) = A(t-d)$. Hence, equation (3) may be rewritten as

$$A*\delta_{d\max}(t) \leq D(t) \leq A*\delta_{d\min}(t) \quad (4)$$

Comparing equation (4) with minimum and maximum service curve definition, it could be stated that the delay element is a service curve element with minimum value $\delta_{dmin}$ and maximum value $\delta_{dmax}$. Mathematically,

$$d_{max} = \inf\{d : d \geq 0, E*\delta_{d\max} \leq S\} \quad (5)$$

and

$$d_{min} = \sup\{t : \overline{s(t)}\} = 0 \quad (6)$$

*B. Buffer requirement*

Buffer requirement is another important parameter to adjust for providing guaranteed service to end users. Too large buffer space introduces high end to end delay. For traffic with high temporal dependencies, such long delay results in data delivery which is of no use to the end users. These data unnecessarily consume network bandwidth. If buffer space is too small more number of packets is dropped. In case of stream traffic, around 2-5% of packet drop is acceptable to end users [1]. But for elastic traffic even a single packet loss is unacceptable and needs retransmission of the packet. These retransmitted packets consume considerable amount of bandwidth. In this subsection buffer requirement for traffic session is analyzed in terms of arrival and departure process and service curve. Maximum buffer requirement may be expressed as,

$$A(t) - D(t) \leq b_{max} \quad (7)$$

i.e. when amount of data arrived is larger than the data that could be departed from the system than backlog is generated. In terms of envelop E and service curve S

$$E(t) \leq b_{max} + S(t) \quad (8)$$

From the definition of envelop and service curve

$$A(t) - D(t) \leq A*(b_{max} + S(t)) - A*S(t)$$
$$A(t) - D(t) \leq A*b_{max} + A*S(t) - A*S(t)$$
$$A(t) - D(t) \leq A*b_{max}$$
$$\text{Or } b_{max} = \sup\{E(t) - S(t) : t \in \Re\} \quad (9)$$

*C. Link Capacity*

Every physical device in the network is connected via communication link of fixed capacity which is known as topology. The maximum transmission capacity or simply the capacity of a link is measured in bits per second or bps. This link capacity also has direct influence on the performance of the network and hence on the quality of service to be offered to a particular flow. So link capacity can also be modeled as a service curve element with maximum allowable transmission capacity in terms of bps.

Let us assume that the capacity of the link under observation is *C* bps. It implies that for any instance of time *t*, it can transmit only *C(t)* amount of data through it. In terms of departure process *D(t)*, it holds the inequality,

$$D(t) \leq C(t)$$

also for any $\tau \leq T$, we can have

$$D(t) - D(\tau) \leq C(t - \tau)$$

Since we have mentioned that $D(t) \leq A(t)$, so

$$D(t) \leq A(\tau) + C(t - \tau)$$

or

$$D \leq A * C \quad (12)$$

In the above expression, $C$ is the amount of data which is passed through the link and denoted by $Ct$ for $t \geq 0$ and $0$ otherwise. It is worthwhile to mention that when a single link is shared by number of different sessions, then the total capacity of the link is not dedicated to a single flow. To provide QoS to end users an amount of the capacity needs to be reserved in advance.

For the arrival process, $A(t)$ passes through the link with lower service curve $S(t)=Ct$, we can find the minimum link capacity required to suffer a maximum delay of $T$ by any traffic flow [12]. Source generating data at a rate $r$, has envelop $E = rt$. If the flow is regulated through $(\sigma,\rho)$ process, the envelop $E(t)=min(rt,(\sigma+\rho t))$. Equation (5) may be rewritten as

$$d_{max} = \inf\{d : E * \delta_d \leq S\} \quad (13)$$

This may be shown graphically as below [12],

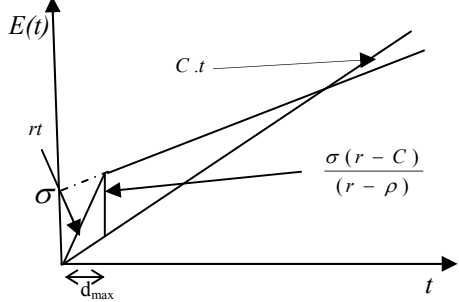

Figure 3: Representation of maximum delay and buffer size

Depending on $d_{max}$ the envelop may fall below or exceed service curve $S(t)$. To minimize the backlog in the system, $d_{max}$ should be shifted to left such that it falls below $S(t)$. In such case,

$$b_{max} = \frac{\sigma(r - C)}{C(r - \rho)} \quad (14)$$

If $r > \rho$, link capacity decreases and $b_{max}$ changes from 0 to $\frac{\sigma}{C}$. If $C < \rho$, it leads to infinite buffer build up. In terms of Envelop and $A(t)$

$$d_{max} = \sup_{t \geq 0}\left(\frac{E(t) - Ct}{C}\right) \quad (15)$$

A network session which allow maximum delay of T,

$$\sup_{t \geq 0}\left(\frac{E(t) - Ct}{C}\right) \leq T$$

Here $C = C_{min}$, and hence

$$\sup_{t \geq 0}\left(\frac{E(t) - C_{min}\, t}{C_{min}}\right) \leq T$$

$$E(t) - C_{min}\, t \leq C_{min} T$$

Or

$$C_{min} \geq \sup_{t \geq 0}\left(\frac{E(t)}{T + t}\right).$$

Strict inequality cannot hold in this case. So,

$$C_{min} = \frac{E(t)}{T + t} \quad (16)$$

VI. RESULTS AND DISCUSSIONS

Results are discussed with respect to wired-cum-wireless scenario depicted in Figure 2. To realize the traffic flow we have considered that several traffic sessions exist from a mobile node (MN) to a correspondent node (CN) via the Access Router (AR). Each of the MN may have multiple number of ongoing sessions and maintains an envelope of $(\sigma, \rho)$ where $\sigma=5$kb and $\rho=200$kbps. Different parameters like end-to-end delay, buffer requirements and link capacity are observed for various wireless link capacities of 64kbps, 128kbps and 256kbps through which MNs are connected to the AR with a propagation delay of 20ms. AR serves incoming packets at a rate of 400kbps and takes a processing delay of 10ms. AR is connected to the CN via a wired T1 link of capacity 1.2mbps and packet suffers a propagation delay of 25ms. To simplify our analysis we have segmented the entire architecture of Figure 2 into two parts: wireless and wired and examined the parameters separately. Then we combined these parameters to evaluate the resources required to provide guaranteed service for the whole architecture.

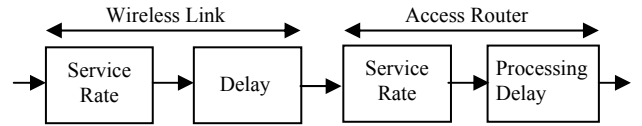

Figure 4: Service curve elements in Wireless segment

With this example scenario, now we will examine the performance of the network in terms of service curve elements end-to-end delay and buffer requirement and bandwidth required so that a guaranteed service could be provided by reserving such resources. Stated scenario may be represented by a series of service curve element as given in figure 4.

*E(t)* : envelope of the incoming traffic
*$S_1(t)$*: service rate of wireless link
*$S_2(t)$*: propagation delay suffered by packet in the wireless segment

$S_3(t)$: rate of service offered by AR
$S_4(t)$: service rate of wired link
$S_5(t)$: propagation delay suffered in wired link

$$E(t) = 5kb + 200\,kbps\,.t \quad (17)$$
$$s_1(t) = 64\,kbps\,.t \quad (18)$$
$$s_2(t) = \delta_{20\,ms}(t) = 20\,ms\,.t \quad (19)$$
$$s_3(t) = -0.4kb + 400\,kbps\,.t \quad (20)$$
$$s_4(t) = 1.2\,mbps\,.t \quad (21)$$
$$s_5(t) = \delta_{25\,ms}(t) = 25\,ms\,.t \quad (22)$$

Based on discussion in section V we have represented the characteristics of different elements of example parameters in terms of service curve. Inclusion of $t$ in all the equations is to represent them as causal process of time. Equation (17) represents the envelop maintained by the source with parameters $\sigma=5kb$ and $\rho=200kbps$ and only the rate ($\rho$) of injecting data in to the network is dependent on the time. Equation (18) and (20) is the service curve of wireless and wired link capacity respectively, whereas, the delay for both the links are represented by equation (19) and (22). Equation (22) models the service of AR. All though AR has a service rate of 400 kbps, in reality actual service rate is lower than 400kbps because of the processing time of the AR. The negative component of the equation (22) reflects the processing delay of the AR in is service curve.

Now from the service curve representation of individual network elements, the maximum and minimum service curve for the wireless segment is represented by equation (23) and (25) respectively. Substituting values for parameters in the equations (25) and (26) and expanding as in [2], equation (24) and (26) is derived.

$$\overline{s} = s_1 * s_2 \quad (23)$$

and calculated as,

$$\overline{s} = -72.8kb + 64\,kb/s.t \quad (24)$$

Similarly, minimum service curve is given by

$$s = s_1 * s_2 * s_3 \quad (25)$$

and calculated as

$$s = -26.4kb + 64\,kb/s.t \quad (26)$$

Now, we will make a measure of different resources required for providing guaranteed service to end users [12,2,3]. Consider that $\theta_i, \phi_i, r_i$ and $p_i$ are the propagation delay, processing delay, transmission rate and service rate respectively at i-th service curve element. So, from $S_1(t)$, $r = \min_{1 \leq i \leq k} r_i$ is the minimum service rate of the wireless segment. Again, from $S_2(t)$, $\theta = \sum_{i=1}^{k} \theta_i = 20ms$ where $k$ is the number of service curve elements that introduces propagation delay in the wireless network. Similarly, the total processing delay in the system is for the service curve element $S_3(t)$ and it is denoted by $\phi = \sum_{i=1}^{k} \phi_i = 10ms$ where $k$ is the number of service curve elements that introduces processing delay in the network. Using envelop of the source $\sigma=5kb$ and $\rho=200kbps$. There is only one router (AR) that forwards the packet with a service rate of 400 kbps and it is the total service rate of the system denoted by $p$, i.e. $p=400$ kbps. From the above discussions the following parameters are derived and given in Table -I:

Minimum delay = $\theta$=20 ms,
Maximum delay = $\theta+\sigma/r$=20ms+12.5ms=32.5ms
Maximum buffer size = $\sigma+\rho\theta$=5kb+9kb=14kb

TABLE I :: DELAY AND BUFFER REQUIREMENT

| Wireless Bandwidth (kbps) | Minimum Delay (ms) | Maximum Delay (ms) | Buffer Size (kb) |
|---|---|---|---|
| 64 | 20 | 32.5078125 | 14 |
| 128 | 14 | 38.039062 | 56 |
| 256 | 12 | 35.019531 | 100 |

Consider an example of a voice coder emitting *10ms* speech and *10 bytes* per frame. The rate of injecting data to the channel is *8kbps*. In the Table-II, we have shown the minimum bandwidth required by the application and number of such sessions that can exhaust the link capacity.

TABLE II : DELAY FOR VOICE DATA

| Wireless Bandwidth (kbps) | Acceptable Delay (ms) | Buffer Space Required (kb) | Sessions Supported |
|---|---|---|---|
| 64 | 35.078125 | 130 | 8 |
| 128 | 35.039062 | 250 | 16 |
| 256 | 35.019531 | 500 | 32 |

Data shown in Table–II represents maximum tolerable delay by each of the session, total session that can exhaust entire bandwidth of a wireless link and the buffer space required to cope up with the delay by the end application. Recorded data shows that maximum end-to-end delay and buffer space required per session decreases when link capacity increases. On the other hand, total sessions supported are more if channel capacity is large. Increased supported session increases the buffer space required to provide guaranteed service at the AR of the wired-cum-wireless scenario. The AR is connected to the destination via T1 link of capacity 1.2mbps. Next follows a calculation of the above observed parameters for wired part of the network. Assumption is made that AR does not maintain any regulator and allows passing data as and when channel is free. For the AR, the following parameters are assumed

φ: processing delay of AR = 10ms

C : capacity of outgoing T1 link = 1.2 Mbps
$L_{max}$ : size of a packet = 8kb

$$\text{Propagation Delay} = \frac{L_{max}}{C} = \frac{8\,kbps}{1.2\,Mbps} = 6.67\,ms$$

It adds another delay to the outgoing data packets as processing delay of AR and propagation delay of wired link (*10ms and 6.67ms*). The link capacity consumed *($C_{con}$)* by the session is calculated as

$$C_{con} = \frac{C}{L_{off}}$$

where $L_{off}$ is the offered load and C is the link capacity. Applying data from Table II and assuming there are sufficiently large number of MNs available to exhaust the total capacity of the wireless link during transmission of data to AR. For the first case with wireless link capacity *64kbps*, there are eight numbers of sessions that continuously sending data to the AR. So, number of sessions that can exhaust the entire wired bandwidth is calculated as,

$$C_{con} = \frac{1.2\,Mbps}{64\,Kbps} = 18.75 \approx 18$$

Approximately 18 numbers of MNs can be supported by the AR distributed over various parts of the cell offering eight sessions per MN. Similarly, in Table – III, we have noted down the number of supported MNs for wireless link of various capacity.

TABLE III :: MN SUPPORTED IN THE SCENARIO

| Wireless Bandwidth (Kbps) | Number of MNs supported | Session supported |
|---|---|---|
| 64 | 18 | 8 |
| 128 | 9 | 16 |
| 256 | 4 | 32 |

Data given in the table-III provides a measure of various parameters like total allowable sessions per mobile node in a sample wired-cum-wireless scenario. In such scenario we have measured the buffer requirement and link capacity required to meet a pre-specified end-to-end delay. The measured quantities can be used to provide guaranteed service to end-users by reserving different resources before starting the conversation. We can make use of the well known Resource Reservation Protocol (RSVP) [10] of Integrated services (IntServ) for reserving resources. The calculations and data presented here provide a mechanism to use network calculus to compute resource requirements to apply Integrated Services in wired-cum-wireless environment where resources are very precious.

VII. CONCLUSION

In this paper, an analysis is made for different sessions in a wired cum wireless environment. The main aim of this paper is to measure end-to-end delay suffered by each of the flow in such scenario and buffer requirement to cope up with the end-to-end delay and minimum bandwidth required by end applications so that a guaranteed service can be provided to the end user. All the service curve elements that contributed to the flow are identified first and their influences to maintain a minimum and maximum service curve is analyzed. With the help of examples the parameters like maximum and minimum delay suffered by a packet and the total buffer requirement is also shown in this paper. The process can be extended to any number of nodes related to the wireless environments. With the growing world of wireless users in this day, guaranteed QoS is a major concern. This paper shows a way to calculate the QoS parameters in a wired-cum-wireless scenario.